# Germanium electrostatic quantum dot with integrated charge detector in an MOS structure

**G. Mazzeo [1,a], E. Yablonovitch [2], H. W. Jiang [3]**

[1] Department of Electrical Engineering, UCLA, Los Angeles, CA 90095

[2] Department of Electrical Engineering, University of California at Berkeley, Berkeley, CA 94720

[3] Department of Physics and Astronomy, UCLA, Los Angeles, CA 90095

ABSTRACT

We report the fabrication and characterization of an electrostatic quantum dot in pure Germanium with an integrated charge measurement transistor. The device uses the $Al_2O_3$/Germanium interface for the confinement of carriers in the Germanium and an hybrid design with an electron quantum dot and hole transistor for the charge detection. The hole transistor, using with NiGe source and drain contacts, despite the modest low temperature carrier mobility of 450 $cm^2$/Vs, has shown a sensitivity to the dot electric potential sufficient to detect single charges tunneling in and out of the quantum dot. The device is realized with a two level gate stack, with the top level used to attract electrons and the lower one to define the electron confinement potential and accumulate the hole transistor. The possibility to improve the device operation using an $Al_2O_3$/SiGe/Ge multilayer for the confinement of electrons at a smoother interface is discussed.

[a] G. Mazzeo is now with MDM-IMM-CNR National Lab, Agrate Brianza, Italy

## *Introduction*

Electrons in Silicon and Germanium share the possibility to reach long spin coherence time enabled by the natural abundance of spinless nuclei and the possibility of isotope purification. While research is now focused on Silicon devices for technological advantages of this material system, Germanium based devices would have properties already recognized in the past[1] that would solve some issues currently open in Silicon devices, so that the initial development efforts would be justified. The use of Germanium rich heterostructures was initially proposed for the peculiar possibilities allowed by conduction band engineering. Bulk Silicon, Germanium and any alloy of these two materials have two local conduction band minima in the $\Delta$ and in the L crystallographic direction. Electrons in either minimum have properties, such as effective mass and Landé *g*-factor, substantially different, but such values are only marginally dependent on the alloy composition: as a consequence in any SiGe substrate or thin film the electronic properties are set by which conduction band local minimum is at the absolute minimum energy [2]. The band profile is affected not only by the alloy composition, but also by the strain involved in the growth of Silicon-Germanium heterostructures. The unique property found in germanium rich heterostructures, and in particular in SiGe films grown on pure Germanium substrates, is that when the Silicon content in the epilayer is larger than 15 or 30% (depending if (100) or (111) substrate orientation is considered) the strain pushes the energy of the X valley at the absolute energy minimum [1,3]. Consequently thin films where either the L or the X valley is the absolute minimum, and thus where electrons can have very different physical properties, can be grown on each other. Besides, the L valley minimum, that is the absolute minimum in Germanium like materials, has advantageous properties deriving from its symmetry properties. This minimum has a 4-fold degeneracy in the bulk, with the energy ellipsoids aligned along the (111) directions. Owing to the fact that the effective mass is larger in the (111) direction, the degeneracy is completely lifted in a 2D confinement structure, provided that the confinement is obtained in a plane normal to the (111) direction: this is in contrast to what happens in Silicon based devices where a 2-fold degeneracy is left. Besides the in-plane mass in Germanium is smaller than in Silicon ($0.082\times m_0$ vs $0.19\times m_0$ for Si and $0.063\times m_0$ for GaAs) allowing a larger quantum dot size, and consequently relaxing the technological requirements.

A type I alignment of the conduction and valence bands is predicted at the SiGe/Ge interface and a 55meV conduction band offset has been recently measured at the $Si_{0.11}Ge_{0.89}$/Ge interface on a (100) substrate [3,4]. In a similar band structure it would be in principle possible to confine electrons in the Germanium substrate, but the limited barrier height requires a more complex device design. The semiconductor/dielectric interface can be a valid alternative to confine the charge carriers and quantum dot devices based on MOS structures are today being successfully realized in Silicon [5,6]. The improvements reported on the Germanium/high-*k* dielectric interface [7] suggest the possibility to transfer the Silicon device design to Germanium. However the operation at very low temperature poses new difficulties that haven't been addressed in Germanium MOS devices yet, namely the realization of low temperature MOSFETs. The integration of a transistor functioning as a highly sensitive electrometer is necessary to detect the charge confined in the quantum dot, and perform the readout of the spin state through projective measurements that have been largely proven in GaAs [8] and recently in Silicon [9] based devices. The operation of MOS devices should not in principle be modified at low temperature, however the freeze-out of the dopants in non degenerately doped source and drain wells and the presence of energy barriers from the S-D metal to the transistor channel, that may be shallow enough to not hinder the operation at room temperature, can easily lead to failure when the device is cooled down to liquid helium temperature or lower. Virtually every metal on Germanium shows a Fermi level pinned close to the valence band [10] rendering necessary a very large doping to enable carrier tunneling through the Schottky barrier. At the same time dopants in Germanium show large diffusivity and lower solid solubility [11] compared to Silicon, making it difficult to reach the required high doping density. Methods suggested recently to reduce the effective barrier height at the metal-semiconductor interface [12,13] didn't provide the needed improvement for operation at 4 K or below, so that failure of *n*-type MOSFETs realized during the initial phases of this work was routinely observed. On the other hand the alignment of the metal level close to the Germanium valence band is favorable to the injection of holes and thus the formation of *p*-MOS devices. The possibility to use hole, rather than electron, qubits has been recently studied [14,15,16] as a way to reduce the influence of the nuclear spins on the spin coherence time, in particular in AlGaAs based devices, using the peculiarity of the hole wavefunction that goes to zero at the nuclei sites. This wouldn’t however be advantageous in Germanium and Silicon based devices, where isotope purification allows the complete elimination of nuclear spins. For group IV materials electrons still represent the preferred choice for the quantum information storage, allowing the use of ESR for spin rotation, having a lower in-plane mass and possibly achieving longer coherence time in semiconductors with spinless nuclei. Consequently we designed and realized a hybrid device where a *p*-type transistor is electrostatically coupled to an electron quantum dot to detect the confined charge.

***Device design***

The device here reported is an MOS structure with a 2 level gate stack. The device architecture already used in Silicon MOS quantum dots [5] has been adapted for the use of carriers of opposite sign in the quantum dot and the measurement channel. The mid level gate, besides depleting the 2DEG and forming the quantum dot, is used to accumulate the transistor channel, while the top level gate is positively biased to attract electrons. The transistor gate is shaped with a small constriction (W/L= 70nm/100nm) constituting the high sensitivity region. Next to this a set of 4 gates, represented in Fig 1c, is used to shape the electron potential well. The two gates indicated as $V_b$ form an adjustable tunnel barrier to the electron reservoir, while the $V_p$ gates are used to tune the energy level in the quantum dot. The entire area represented in figure 1c is covered by the positively biased top gate, not reproduced in the picture, used to attract the electrons.

To optimize and validate the gates design we simulated the device operation with a procedure similar to what described in [17]. In Fig 1d we report the surface charge density calculated when both the hole channel and the electrons in the quantum dot are accumulated. Besides showing the formation a quantum dot for electrons closely coupled to the hole transistor this picture evidences the formation, in the area where no mid-level gates are present, of a 2DEG acting as an electron reservoir for the quantum dot. An *n*-type ohmic contact used to inject electrons in

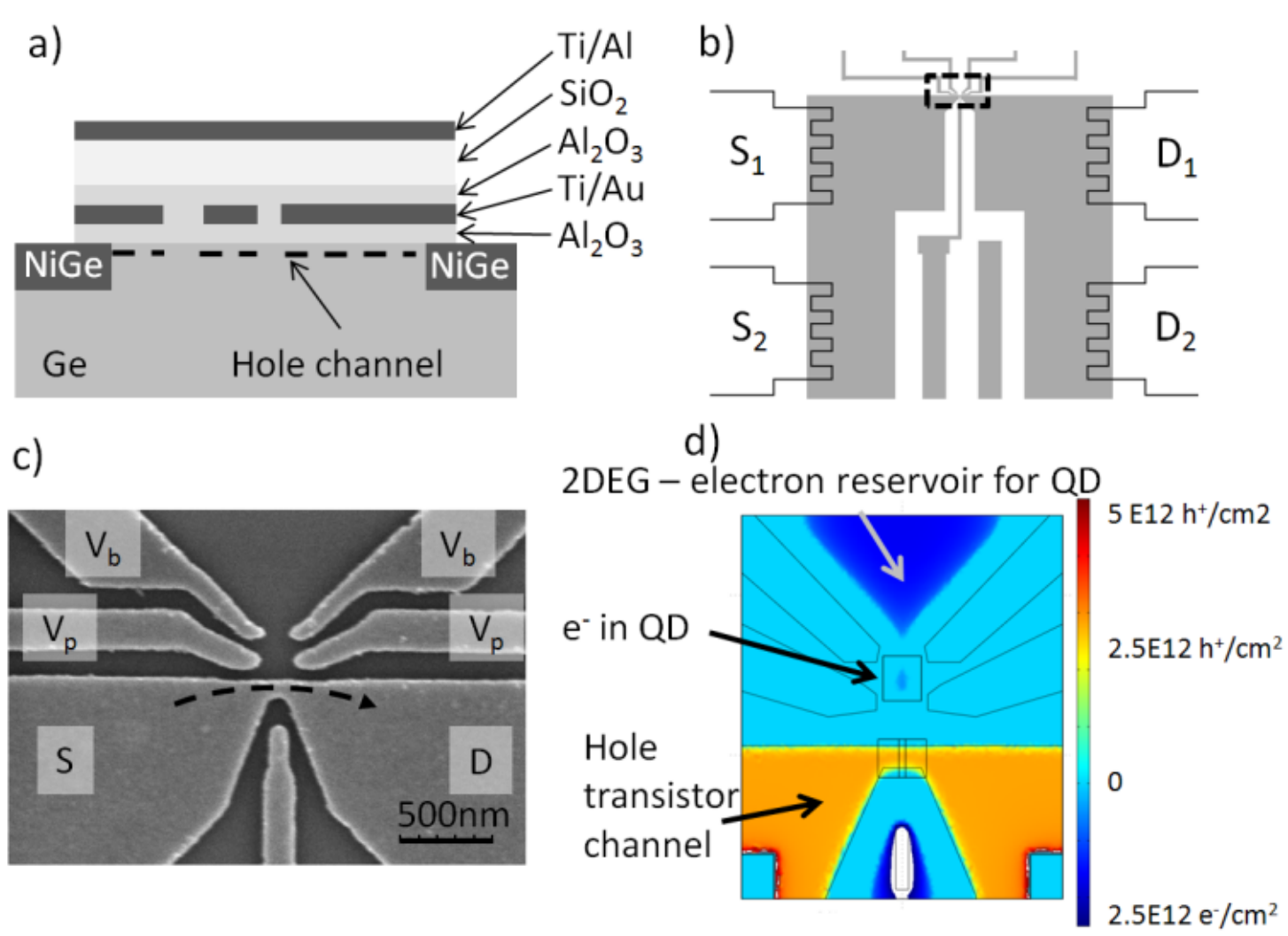

*Fig1: a) Schematic cross-section of the device here reported. b) Gate structure of the quantum dot device, with S1, S2, D1, D2 indicating the ohmic contacts, and the gray area the mid-level gates. The top gate covers the entire device and is not represented. c) SEM image of the device in the area delimited by the dashed square in fig 1b. The arrow indicates the region where the hole current flows. d) Simulated surface charge density at typical operation condition for the formation of an electron quantum dot*

this 2DEG would have less stringent requirements compared to the source and drain contacts of the transistors, so that the technological difficulties previously discussed wouldn't play a major role: indeed during the operation electrons are exchanged between the QD and the reservoir, leaving the total accumulated charge in the 2DEG constant. Even a very large contact resistance for the *n*-type contact would thus be tolerable: for example a $10^{12}$ $\Omega$ contact resistance would still allow accumulating the 2DEG over a 100 x 100 $\mu m^2$ area device in tens of milliseconds. Besides, even if no ohmic contacts are available, other non ideal effects such as interface defects or band-to-band tunneling may favor the promotion of electrons in conduction band, and more reliable results would be obtained accumulating the 2DEG at room temperature before cool down or generating the electrons by light absorption.

***Fabrication and characterization***

The device previously described was realized on a (100) non-intentionally doped germanium substrate. After standard cleaning procedure, Nickel was deposited on the ohmic contact regions and NiGe was formed via annealing at 400 °C for 4 min in $N_2$ atmosphere. Next we deposited the first gate layer consisting of 44 nm $Al_2O_3$ by atomic layer deposition at 200 °C. The mid level gates were defined by e-beam lithography and deposition of Ti/Au. The second gate dielectric was a stack of 44nm $Al_2O_3$ and 100nm $SiO_2$. The extra $SiO_2$ layer was added to reduce the capacitance of quantum dot to the top gate, increasing the charging energy and the electrometer sensitivity. Finally the top gate was defined by optical lithography and deposition of Ti/Al.

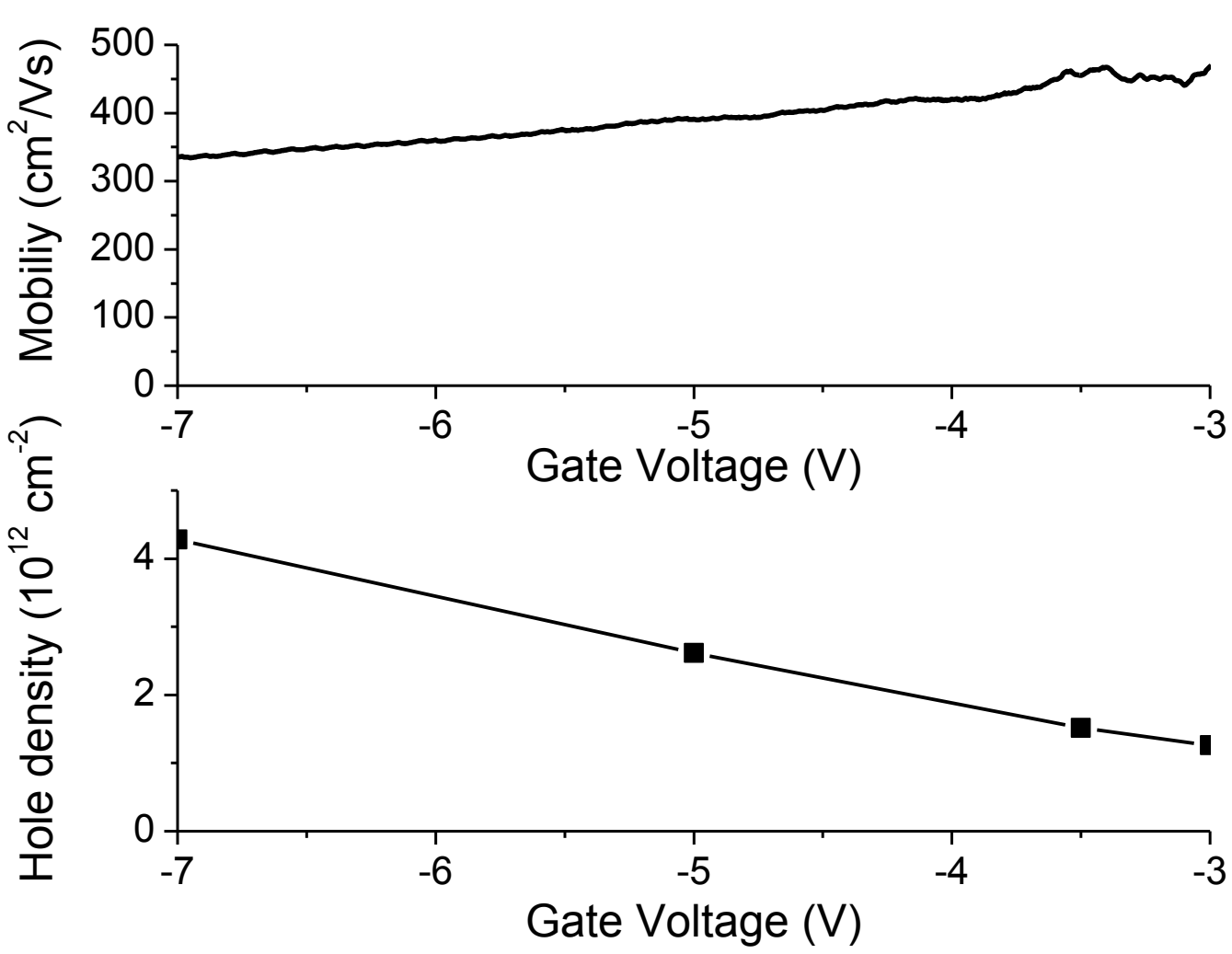

*Fig 2: Hole mobility and density measured as a function of the gate voltage at 0.37 K in a gated Hall bar device*

Device testing was performed in a $^3$He refrigerator at a base temperature of 0.37K. The performances of the hole transistor were first measured on a gated Hall bar device. At the base temperature we found that the source and drain contacts consistently maintained their ohmic behavior. The curve of Hall density vs. gate bias shows a good linear behavior. The minimum density we could measure on Hall bar was $1.2\times10^{12}$ $h^+/cm^2$ that is evidence of a large trapping of carriers not contributing to the conduction. We also observe an increase of the threshold voltage from 4.2 to 0.37 K, further evidencing the presence of a large density of non mobile carriers. Finally, the mobility vs. gate voltage curve, reported in fig. 2, showed a peak value of 450 $cm^2/Vs$. These evidences show that the 2DHG properties are mainly influenced by the interface scattering: this is not surprising considering that our dielectric consists of $Al_2O_3$ deposited on the free germanium surface without use of interface layers or in situ surface cleaning.

Moving on to the properties of the charge detector, reported in figure 3a, even though the ohmic contacts constantly showed ohmic characteristics, the drain current vs. gate voltage ($I_d$-$V_g$) curves on this device didn't present a smooth linear behavior close to threshold, but several peaks and valley characteristic of the Coulomb blockade regime that evolved to a linear characteristic when the gate was made more negative. While the transistor gate was not

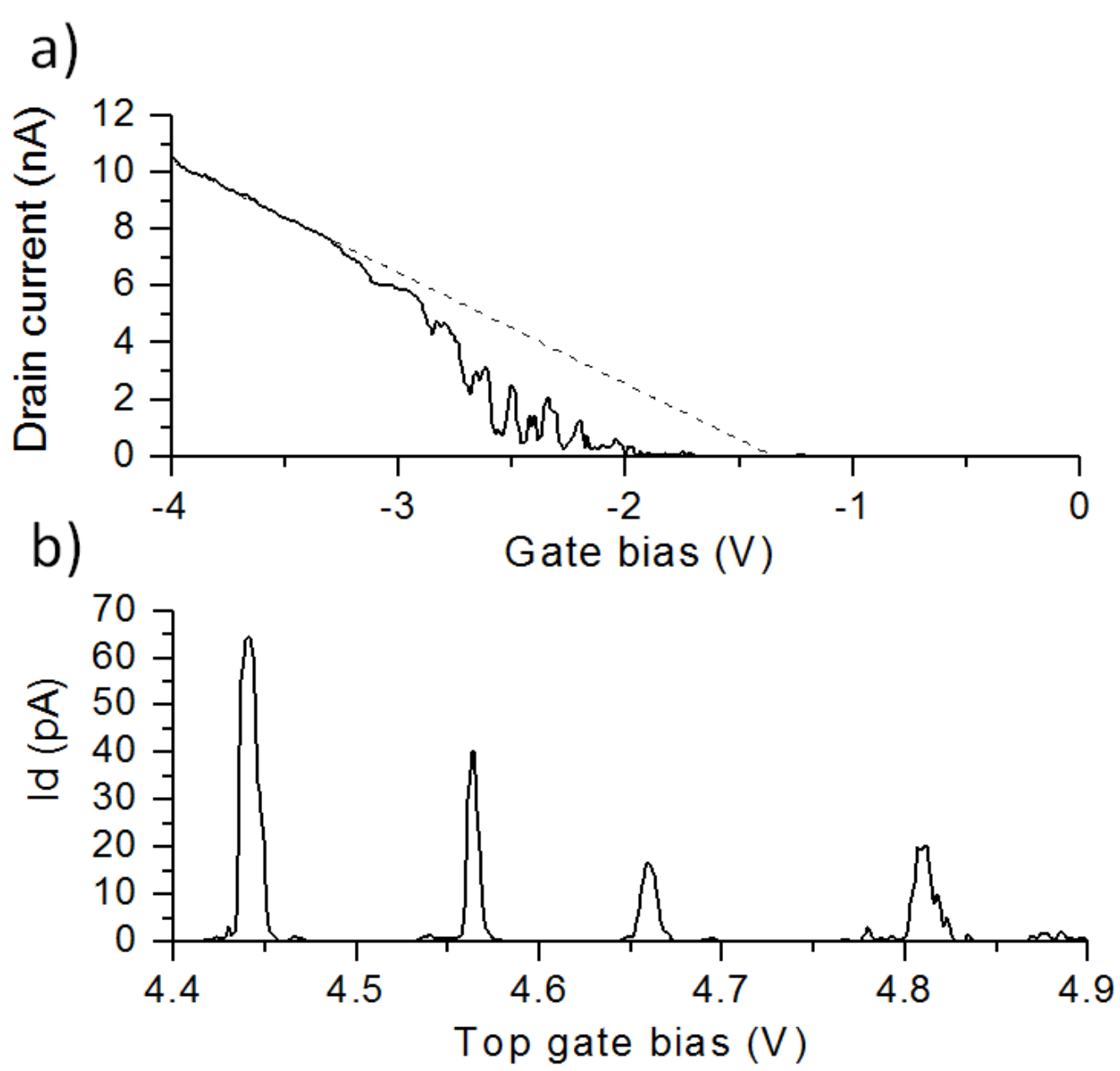

*Fig3: a) Source-Drain current vs. Gate voltage through the charge measurement transistor evidencing a coulomb blockade regime at low gate overdrive. The straight line is computed with the threshold voltages measured on the Hall bar, a mobility of 450 cm$^2$/Vs and the geometric aspect ratio of the transistor gate. b) Id vs. top gate voltage measure near threshold showing a clear single hole transistor behavior.*

shaped to form a quantum dot, this characteristic is not surprising, as potential fluctuations at the $Al_2O_3$/Ge interface caused from interface defects or charges in the oxide can lead to the formation of isolated potential minima that can be screened when a sufficient density of holes is accumulated. Finer I-V curves, reported in Fig 3b, measured close to threshold confirmed a behavior typical of a quantum dot in the Coulomb Blockade regime. In this case the channel potential is controlled via the top gate: considering that only in the region of the constriction the field lines from the top gate can penetrate under the transistor gate, we conclude that the non linearity is due to the transport through the small constriction rather than to the contacts or the leads.

Tuning the transistor in the quantum dot regime would allow the maximum sensitivity to the neighboring charges. However, as evidenced by both 2-gate scans and by stability diagrams, not reported here, the conduction is obtained through a random distribution of multiple quantum dots: in such condition is was hard to achieve the needed stability of the sensitivity. Instead we worked at larger bias where a smoother and stable $I_d$-$V_g$ characteristic was obtained. The transistor sensitivity reported in figure 4a, defined as the derivative of the S-D current to the $V_b$ bias and measured at a 1 mV source-drain voltage and 20 $mV_{rms}$ modulation of $V_b$, evidences a still non linear, but reproducible, response of the transistor. The average measured sensitivity is

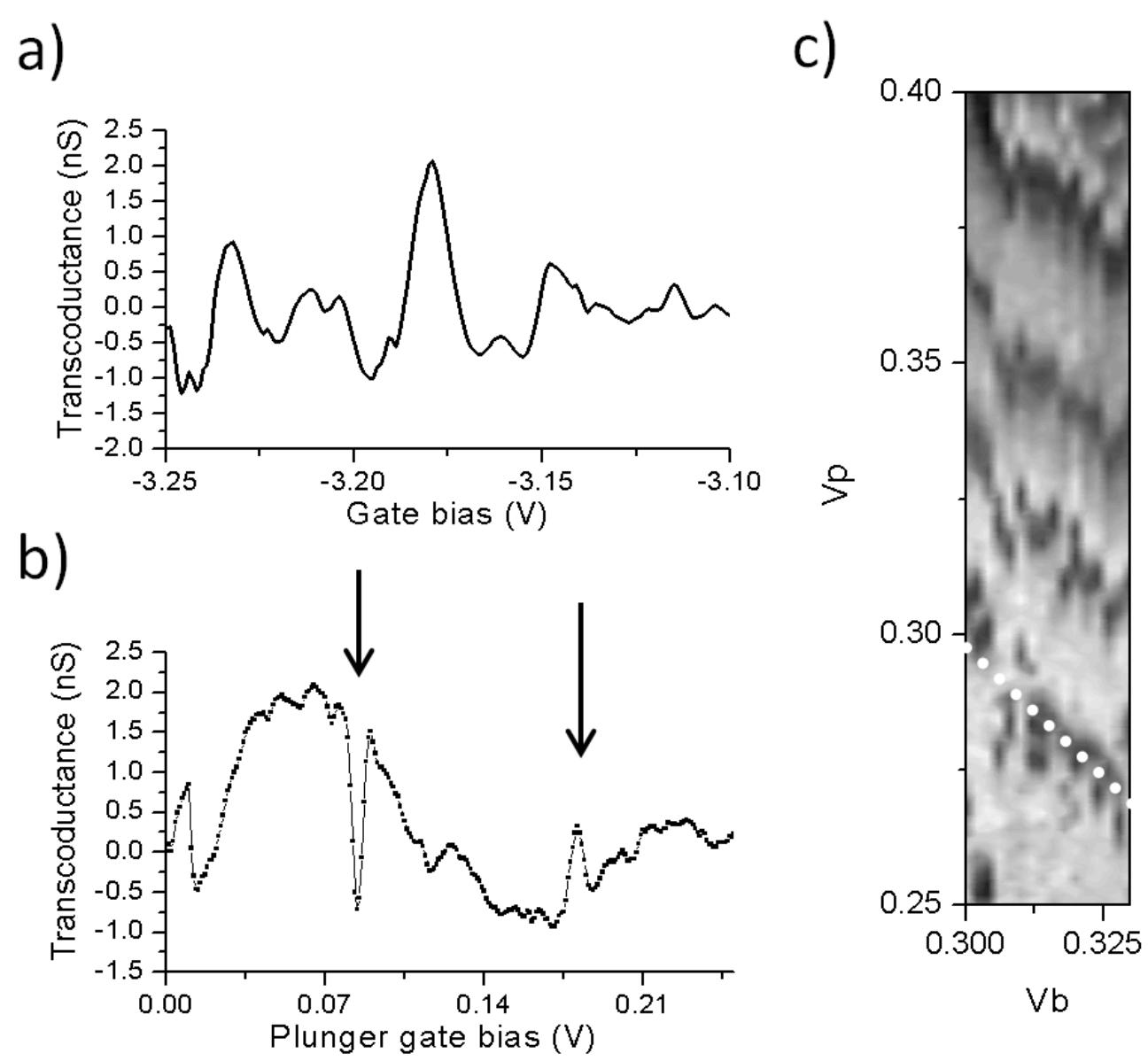

*Fig4: a) transistor sensitivity to the barrier gate bias ($dI_{sd}/dV_b$) as a function of $V_g$. b) transistor sensitivity ($dI_{sd}/dV_b$ ) as a function of $V_p$. The dips marked are evidence of single electrons entering/leaving the dot driven by the modulation of $V_b$ . c) 2D scan of the sensitivity vs. $V_p$ and $V_s$. The dotted line is a eye guide representing a slope of 1V/V, corresponding to and equivalent electrostatic coupling of the electrons to $V_p$ and $V_b$ gates.*

$1.3 \times 10^{-10}$ A/V with peaks up to $3 \times 10^{-9}$ A/V. The sensitivity is about one order of magnitude lower compared to what reported for similar Si based devices [5] and can be attributed to the lower hole mobility that requires higher charge density in the channel to obtain comparable source-drain resistance.

When the plunger gate is swept the same oscillations on the sensitivity curves are observed with a longer period as required from the weaker coupling of this gate to the channel. Superposed to these curves are sharp deeps evidenced in Fig. 4b that are signature of single electrons tunneling in and out of the quantum dot, coherently driven from the modulation of $V_b$. To confirm that the switching events can be attributed to charge in the quantum dot, rather than to defects in other positions of the device, we observe the shift of the peaks as a function of the two quantum dot gates, $V_s$ and $V_b$. These data are reported in Fig. 4c where the dotted line represents the slope of 1 V/V. This slope confirms the comparable coupling strength of the charge to the two set of gates, compatible with the position of the charge inside the quantum dot region.

Integrating the dip in the trace in Fig 4b we can extract the change of source-drain current after the addition/subtraction of one electron to the quantum dot that, in the highest sensitivity region, corresponds to a current step of $1.1 \times 10^{-11}$ A. This is obtained for a total current $I_{sd} = 3.5 \times 10^{-9}$ A and source-drain bias $V_{sd} = 1$ mV. At 0.37 K the main noise source in these conditions is the shot noise, giving a maximum theoretical speed for the detection of single tunneling events of $10^5$ Hz.

The instabilities measured in the 2D scans evidence the presence of trapped charges in the quantum dot, expected given the nature of the *$Ge/Al_2O_3$* interface, and limit the ability to consistently control the charge accumulated in the quantum dot in a limited range of $V_p$ and $V_b$. Much work has been lately focused on the improvement of the Germanium/dielectric interface, but data on the low temperature mobility of electrons at such interfaces is still lacking. As an alternative approach to the improvement of the device we suggest the use of the SiGe/Ge interface for the confinement of electrons. The possibility to confine electrons in $Si_xGe_{1-x}/Ge$ heterostructures with $x \cong 0.1$ is limited by the relatively small energy barrier available, just 55 meV. However electrons in Germanium have a fairly large longitudinal transverse effective mass, which is the relevant one for tunneling through this kind of barrier if the heterostructure is grown on (111) substrates, so that a sufficiently long confinement of electrons can be obtained, provided that the transverse electric field is limited. This requirement means that the hetero-interface is adequate for the confinement of the few electrons in the quantum dot but not for the accumulation of the 2DEG of the reservoir or the 2DHG of the electrometer channel.

A schematic cross-section of a possible device using both the $Al_2O_3$/SiGe and the SiGe/Ge interface, realized on a $Al_2O_3$/SiGe/Ge multilayer, is reported in figure 5: the high energy barrier at the $Al_2O_3$/SiGe interface is used to confine the 2DHG, that doesn't require high interface quality, while the electrons in the quantum dot requiring a smooth and well controlled confinement potential would be at the SiGe/Ge interface. A 2DEG, acting as an electrons reservoir, would be formed at both the $Al_2O_3$/SiGe and at the SiGe/Ge interface. Electrodes are provided so that the more mobile electrons at the lower interface would have a smaller energy barrier to the quantum dot, so that this layer would be the one actually used for the electron reservoir. The quantum dot potential is set with a metal electrode in contact with the SiGe surface, so that electrons eventually tunneling through the barrier can be collected without modifying the potential of the electron trap. In fig 5b we report a simulation of the potential profile. The large transverse mass of electrons in Germanium makes this approach a viable solution. In the design reported the barrier to confine electrons is triangular with an electric field of 2.2 $10^6$ V/m. Through a WKB approximation, using an effective barrier height of 50meV, we can estimate an average confinement time larger than one hour, large enough that it won't be the limit factor on the complexity of the spin manipulation that can be executed in such a device.

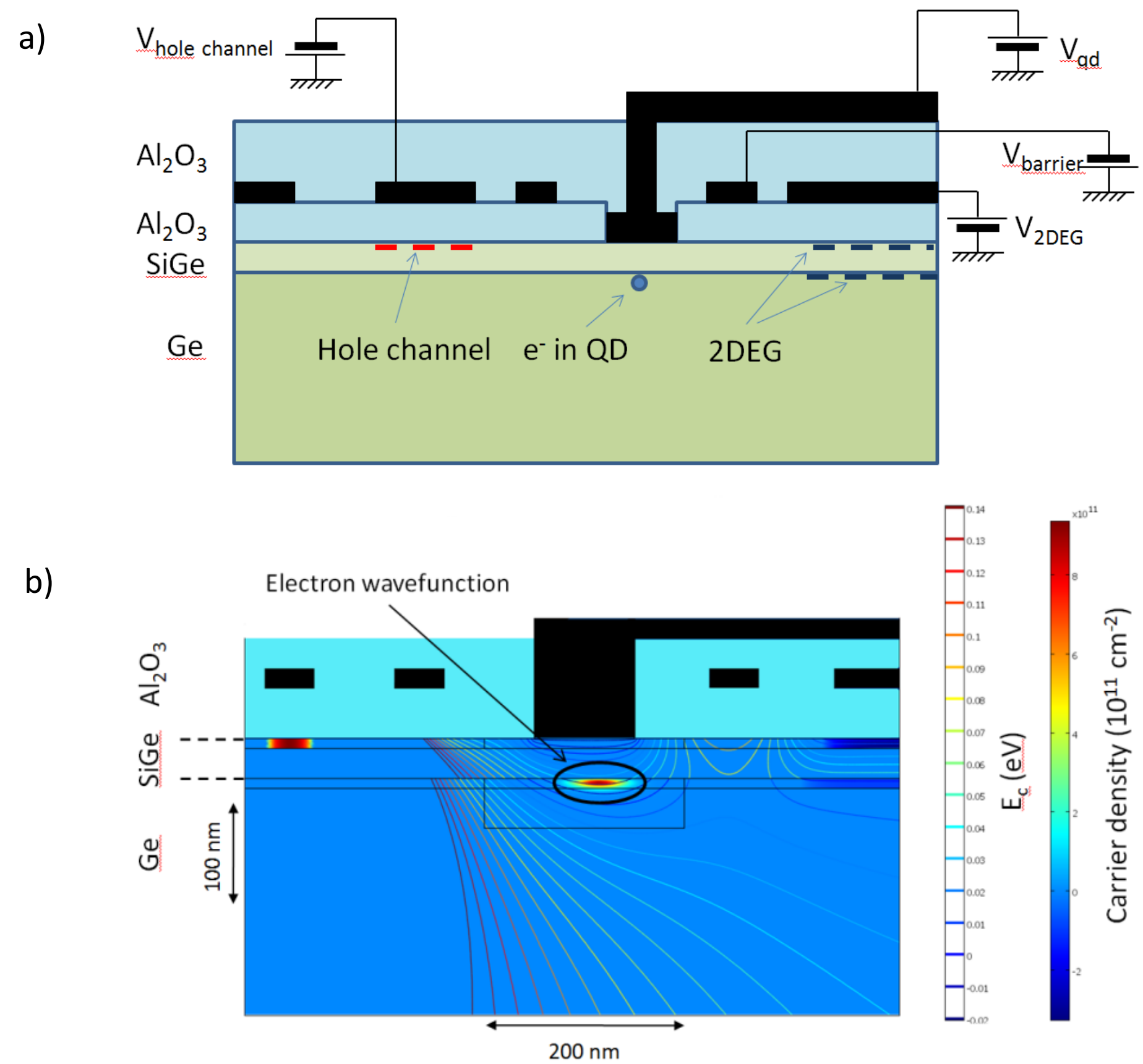

*Fig 5. a) Schematic cross-section of a quantum dot device realized on a SiGe-Ge heterostructure with integrated charge sensor. The 2DHG and the electron in the quantum dot lay on two different planes of the device. The Quantum dot electrode is directly in contact with the SiGe surface in order to collect the electrons possibly tunneling through the barrier. b) Simulation of the device. The color in the Ge and SiGe layers encodes the carrier surface density (positive numbers indicate a hole channel, while negative are referred to electrons) while the iso-curves indicate the conduction band energy, whit 10 meV steps. The electron wavefunction is obtained solving the Schrödinger equation in the potential well formed in the quantum dot region.*

## ***Conclusions***

Concluding, we have shown for the first time the formation of an electron quantum dot in pure Germanium using an $Al_2O_3$/Ge MOS structure with an integrated readout channel. We designed a device employing a *p*-type transistor to detect the electron charge in the quantum dot: this design has allowed overcoming the technological difficulties encountered in the realization of low temperature Germanium *n*-MOSFETs and added no complexity to the device fabrication.

While the low hole mobility of 450 $cm^2$/Vs, caused by the use of the $Al_2O_3$/Ge interface, reduced the transistor current and sensitivity, we were able to detect the displacement of single electrons charges driven by the modulation of one of the quantum dot gates. Control of the quantum dot potential with the two sets of gates confirmed that the charge detected was confined in the quantum dot defined by the electrodes. Instability observed in the 2-gates scans suggest that charges are trapped at the $Al_2O_3$/Ge interface causing non repeatable modifications of the quantum dot potential during the experiments, limiting the control on the quantum dot charge. The low interface quality can be acceptable for the charge detection transistor, where defects can be screened by a larger carrier density and the penalty on the mobility and sensitivity doesn't hinder the detection of single charges. However the electrons in the quantum dot will require a cleaner interface to allow better control of the quantum dot potential: we suggest the use of a three layer structure consisting of $Al_2O_3$/SiGe/Ge. The large barrier at the $Al_2O_3$/SiGe interface is used for the accumulation of the charge measurement transistor, while the electrons that are accumulated with a lower surface density are confined at the SiGe/Ge interface. A proper device design can limit the electric field in the SiGe barrier over the quantum dot, ensuring a confinement time much longer than the time required for electron spin manipulations.`

[1] R. Vrijen, E. Yablonovitch, K. Wang, Kang , H.W. Jiang, W. Hong, A. Balandin V. Roychowdhury, T. Mor, D. DiVincenzo, Electron-spin-resonance transistors for quantum computing in silicon-germanium heterostructures, Phys. Rev. A 62, 012306(2000)

[2] Kasper, Erich; Lyutovich, Klara Properties of Silicon Germanium and SiGe: Carbon. (pp: 124-147). Institution of Engineering and Technology. Online version available at: http://knovel.com/web/portal/browse/display?_EXT_KNOVEL_DISPLAY_bookid=1138&VerticalID=0

[3] M. M. Rieger, P. Vogl, Electronic-band parameters in strained $Si_{1-x}Ge_x$ alloys on $Si_{1-y}Ge_y$ substrates, Phys. Rev. B 48 14276(1993)

[4] G. Mazzeo and E. Yablonovitch and H. W. Jiang and Y. Bai and E. A. Fitzgerald, Conduction band discontinuity and electron confinement at the Si[sub x]Ge[sub 1 - x]/Ge interface, Appl. Phys. Lett. 96 ,213501 (2010).

[5] Xiao, Ming and House, Matt and Jiang, Hong Wen, Measurement of the Spin Relaxation Time of Single Electrons in a Silicon MOS-Based Quantum Dot, arXiv:0909.2857v1 (2009)

[6] E. P. Nordberg, G. A. Ten Eyck, H. L. Stalford, R. P. Muller, R. W. Young, K. Eng, L. A. Tracy, K. D. Childs, J. R. Wendt, R. K. Grubbs, J. Stevens, M. P. Lilly, M. A. Eriksson, and M. S. Carroll, Enhancement-mode double-top-gated metal-oxide-semiconductor nanostructures with tunable lateral geometry, Phys. Rev. B 80, 115331 (2009)

[7] S.J. Lee, C. Zhu, D.L. Kwong, Interface Engineering for High-k Ge MOSFETs, in Advanced Gate Stacks for High-Mobility Semiconductors, Dimoulas, A.; Gusev, E.; McIntyre, P.C.; Heyns, M. (Eds.), Springer Series in Advanced Microelectronics, Vol. 27, pp. 115-138, ISBN: 978-3-540-71490-3

[8] J.M Elzerman, R. Hanson, L. H. Willems van Beveren, B. Witkamp, L. M. K. Vandersypen, L. P. Kouwenhoven, Single-shot read-out of an individual electron spin in a quantum dot, Nature 430, 6998

[9] Andrea Morello, Jarryd J. Pla, Floris A. Zwanenburg, Kok W. Chan, Hans Huebl, Mikko Mottonen, Christopher D. Nugroho, Changyi Yang, Jessica A. van Donkelaar, Andrew D. C. Alves, David N. Jamieson, Christopher C. Escott, Lloyd C. L. Hollenberg, Robert G. Clark, Andrew S. Dzurak, Single-shot readout of an electron spin in silicon, arXiv:1003.2679v3

[10] Tomonori Nishimura,a Koji Kita, and Akira Toriumi, Evidence for strong Fermi-level pinning due to metal-induced gap states at metal/germanium interface, Appl. Phys. Lett 91 123123 (2007)

[11] M. Meuris and B. De Jaeger and J. Van Steenbergen and R. Bonzom and M. Caymax, Germanium Deep-Submicron p -FET and n -FET Devices, Fabricated on Germanium-On-Insulator Substrates, in Advanced Gate Stacks for High-Mobility Semiconductors, pag. 333, vol. 27, Springer Berlin Heidelberg, 2007

12 R. R. Lieten and S. Degroote and M. Kuijk and G. Borghs, Ohmic contact formation on n-type Ge, Appl. Phys. Lett, 92, 022106 (2008)

[13] Yi Zhou and Masaaki Ogawa and Xinhai Han and Kang L. Wang, Alleviation of Fermi-level pinning effect on metal/germanium interface by insertion of an ultrathin aluminum oxide, Applied Physics Letters 93, 202105(2008)

[14] D. Brunner, B. D. Gerardot, P. A. Dalgarno, G. Wüst, K. Karrai, N. G. Stoltz, P. M. Petroff, R. J. Warburton, A Coherent Single-Hole Spin in a Semiconductor, Science, 325, 70(2009)

[15] M. H. Kolodrubetz, J. R. Petta, Coherent Holes in a Semiconductor Quantum Dot, Science, 325 5936 (2009)

[16] D. V. Bulaev and D. Loss, Electric Dipole Spin Resonance for Heavy Holes in Quantum Dots, Phys. Rev. Lett 98 097202 (2007)

[17] Zhang, X. C. and Mazzeo, G.  and Brataas, A.  and Xiao, M.  and Yablonovitch, E.  and Jiang, H. W., Tunable electron counting statistics in a quantum dot at thermal equilibrium, Phys. Rev. B 80, 035321(2009)